\documentclass[9pt,twocolumn,twoside]{osajnl}

\usepackage{amsmath,amssymb}
\usepackage{bm}
\usepackage{float}
\usepackage{multirow}
\usepackage{color}
\newcommand{\eref}[1]{Eq.\hspace{0.5mm}(\ref{#1})}

\newcommand{\fref}[1]{Fig.\hspace{0.5mm}\ref{#1}}
\newcommand{\Fref}[1]{Fig.\hspace{0.5mm}\ref{#1}}
\newcommand{\tref}[1]{Table\hspace{0.5mm}\ref{#1}}

\newcommand{\ms}[0]{\hspace{.25mm}m\hspace{.5mm}s$^{-1}$}
\usepackage{setspace}

\journal{josaa} 

\setboolean{shortarticle}{false} 

\title{Multi time-step wave-front reconstruction for tomographic Adaptive-Optics systems}

\author[1]{Yoshito H. Ono}
\author[1]{Masayuki Akiyama}
\author[2]{Shin Oya}
\author[4]{Olivier Lardière}
\author[4]{David R. Andersen}
\author[5]{Carlos Correia}
\author[6]{Kate Jackson}
\author[3]{Colin Bradley}

\affil[1]{Astronomical Institute, Tohoku University, 6-3 Aramaki, Aoba-ku, Sendai 980-8578, Japan}
\affil[2]{TMT-J Project Office, NAOJ, 2-21-1 Osawa, Mitaka, Tokyo 181-8588, Japan}
\affil[3]{Adaptive Optics Laboratory, University of Victoria, 3800 Finnerty Rd., Victoria V8P 5C2, British Columbia, Canada}
\affil[4]{NRC-Herzberg, 5071 West Saanich Rd., Victoria, British Columbia, Canada}
\affil[5]{Aix Marseille Université, CNRS, LAM (Laboratoire d'Astrophysique de Marseille) UMR 7326, 13388 Marseille, France}
\affil[6]{Division of Engineering and Applied Science, California Institute of Technology, 1200 E. California Boulevard MC 155-44, Pasadena, CA 91125, USA}
\affil[*]{Corresponding author: yo-2007@astr.tohoku.ac.jp}

\dates{Compiled \today}

\ociscodes{(010.1080) Active or adaptive optics; (010.1330) Atmospheric turbulence.}

\doi{xxxxxxx}

\begin{abstract}
In tomographic adaptive-optics (AO) systems, errors due to tomographic wave-front reconstruction limit the performance and angular size of the scientific field of view (FoV), where AO correction is effective. We propose a \textit{multi time-step} tomographic wave-front reconstruction method to reduce the tomographic error by using the measurements from both the current and the previous time-steps simultaneously.  We further outline the method to feed the reconstructor with both wind speed and direction of each turbulence layer. An end-to-end numerical simulation, assuming a multi-object AO (MOAO) system on a 30 m aperture telescope, shows that the multi time-step reconstruction increases the Strehl ratio (SR) over a scientific FoV of 10 arcminutes in diameter by a factor of 1.5--1.8 when compared to the classical tomographic reconstructor, depending on the guide star asterism and with perfect knowledge of wind speeds and directions. We also evaluate the multi time-step reconstruction method and the wind estimation method on the RAVEN demonstrator under laboratory setting conditions. The wind speeds and directions at multiple atmospheric layers are measured successfully in the laboratory experiment by our wind estimation method with errors below 2 \ms. With these wind estimates, the multi time-step reconstructor increases the SR value by a factor of 1.2--1.5, which is consistent with a prediction from end-to-end numerical simulation.
\end{abstract}
\setboolean{displaycopyright}{false}
\begin{document}
\maketitle
\thispagestyle{fancy}
\ifthenelse{\boolean{shortarticle}}{\abscontent}{}
%
\section{INTRODUCTION}
Most large ground-based telescopes have single-conjugate adaptive optics (SCAO) systems \cite{Hayano-10,Wizinowich-06,Lai-14}, which correct phase distortion due to the atmospheric turbulence by using a wavefront sensor (WFS) and a deformable mirror (DM). These systems realize high resolution observations in the near infrared wavelengths. However, the correction of SCAO is effective only within a limited angle from a guide star (GS), known as the isoplanatic angle $\theta_0$. The limitation is caused by the optical path difference of a GS and a science target at high altitude in the atmosphere. The limitation is called angular anisoplanatism. The typical value of $\theta_0$ is roughly 10 arcseconds in \textit{H}-band even at good observing sites \cite{Skidmore-09}.\par
Two wide-field AO (WFAO) concepts are proposed: multi-conjugate AO (MCAO) \cite{Beckers-88} and multi-object AO (MOAO) \cite{Hammer-02}. Both use tomography, which estimates the atmospheric volume to overcome the angular anisoplanatism and to enlarge the size of a scientific field of view (FoV), where the AO correction is effective. These systems use multiple wavefront sensors aiming at GSs in different directions and observe the atmospheric turbulence above a telescope corresponding to the scientific FoV. Then, the three dimensional structure of the phase distortion is estimated from WFS measurements by the tomographic wave-front reconstructor. Both systems use multiple DMs, but in different arrangements, to perform correction. MCAO systems compensate for the phase distortion three-dimensionally by multiple DMs, which are put in series and conjugated at different heights in the atmosphere. MCAO systems provide a uniform correction over a wide corrected FoV, in which the AO correction is performed by DMs. In MOAO systems, there are multiple science pick-off arms on science targets in the wide scientific FoV. Each pick-off arm directs the light from one science target to a DM contained in each science channel. The DM applies the optimal correction to a small corrected FoV in its direction. This parallel approach increases the efficiency of the observation, and can realize, for example, an observation of multiple galaxies with a multi-object integral field spectrogragh fed by MOAO corrected wavefronts \cite{Andersen-06,Hammer-14,Akiyama-14}. MOAO systems require open-loop control because DMs correct the phase distortion in different directions from the WFS and this is challenging in terms of calibration.\par
The diameter corresponding to a given scientific FoV at different altitudes in the atmosphere is called the meta-pupil. It becomes larger at higher altitudes. As a result, there are areas covered by no or only one footprint of a GS optical path in the meta-pupils at high altitudes. The performance of the tomographic reconstruction is affected by these areas. In this paper, we refer the area covered by no or only one footprint of a GS optical path as \textit{uncovered area} or \textit{unoverlapped area}, respectively. The \textit{uncovered area} causes a significant tomographic error because there is no information from the WFS measurements in this region. The \textit{unoverlapped area} also causes a significant tomographic error. The WFS probes a phase distortion integrated in the direction of its GS. In other words, a measurement of one WFS includes phase distortions at multiple altitudes, which causes \textit{degeneracy} in the height direction. The lack of information on the \textit{uncovered} and \textit{unoverlapped} areas can't be fully solved by the tomographic reconstructor leading to a large tomographic error even within the GS asterism.\par
In particular, the tomographic error due to the geometry of the GSs and the atmospheric turbulence becomes more severe when one considers the expanded scientific FoV of a WFAO system with a wide GS asterism. In addition, using laser guide stars (LGS) increases the geometric error compared with using natural guide stars (NGS), due to its conical optical path, which is called the cone effect. This means the geometric error will be more severe in WFAO systems for the future extremely large telescopes (ELT) \cite{Andersen-06,Hammer-14,Akiyama-14}, which have 25$\sim$40 m primary mirror diameters, with multiple LGSs because the cone effect worsens with aperture size.\par
The idea to use wind information for the WFAO control is studied for predictive control, which reduces the lag error resulting from the change of the atmospheric turbulence during the exposure time of the WFS and the computation time for AO corrections \cite{Correia-14,Correia-15}. These predictive controllers allow the use of longer integration times of the WFS and/or real-time processing, and therefore result in an increased limiting magnitude of GSs and improved sky-coverage. In this paper, our method follows a parallel approach: it addresses reducing the tomographic error by increasing numerically the number of GSs, although it remains possible to expand it to include predictive control.\par 
Our reconstruction method is based on Taylor's frozen flow hypothesis according to which the atmospheric turbulence layers move with a constant speed keeping its pattern of phase distortion. Under this assumption, we can estimate the time evolution of the measurements of the WFS at previous time-steps if we know wind speeds and directions, and the previous measurements can be used as the information for the tomographic reconstruction at the current time-step \cite{Ammons-12}. Our idea is to reduce the tomographic error due to the geometry of the GSs and the atmospheric turbulence by increasing the information using the measurements from both the current and the previous time-steps simultaneously and thus, to expand the corrected scientific FoV of WFAO systems. \par
Estimating wind speeds and directions at different altitudes is therefore essential. Several methods to estimate wind speeds and directions from measurements of Shack Hartmann WFS (SH-WFS) are proposed and tested by using on-sky measurements\cite{Poyneer-09,Guesalaga-14}. These methods successfully detect the wind at multiple altitudes and start to be included in the real-time operation. We develop a wind profiler which estimates wind speeds and directions at each altitude by using temporal correlation of the phase distortion pattern reconstructed by the tomographic reconstruction.\par
The multi-step reconstructor is introduced and evaluated analytically in Section 2. In Section 3, we show its optimal performance based on an end-to-end (E2E) numerical simulation, assuming a MOAO system on a 30 m aperture telescope with multiple LGSs. The wind estimation method is outlined in Section 4. In Section 5, we present the performance of the wind estimation method and the tomographic reconstruction evaluated in lab on RAVEN\cite{Olivier-14}, a MOAO technical and science demonstrator installed and tested on the Subaru telescope, and discuss a comparison between the lab-test and a numerical simulation. Finally, we summarize our results in Section 6.
%
\section{TOMOGRAPHIC RECONSTRUCTION METHODS}
%
\subsection{Classical Single Time-Step Tomographic Reconstruction}
Here, we assume that the atmospheric turbulence consists of $N_l$ thin layers located at different altitudes. The aperture-plane phase of the light coming from a GS in the direction $\bm{\theta}=(\theta_x,\theta_y)$ at time $t$ is given by
\begin{equation}\label{eq2A-1}
\bm{\varphi}(\bm{\theta},t)=\sum_{i=1}^{N_l}\bm{P_\theta^i}\bm{\phi_i}(t)=\bm{P_\theta}\bm{\phi}(t)
\end{equation}
where $\bm{\varphi}$ is a column vector of distorted phase values on a discrete grid of points on the telescope aperture, $\bm{\phi_i}(t)$ is a column vector of a phase distortion on a discrete grid on the $i$-th turbulence layers at time $t$, $\bm{\phi}(t)=[\bm{\phi_1}^T(t) \cdots \bm{\phi_{N_l}}^T(t)]^T$, $\bm{P_\theta^i}$ is a ray-tracing submatrix which extracts a phase distortion within a footprint of a GS optical path in the direction $\bm{\theta}$ on the $i$-th atmospheric turbulence by using a bilinear interpolation, and $\bm{P_\theta}$ is a concatenation of all submatrices $\bm{P_\theta^i}$.\par
In MOAO systems, a tomographic reconstructor is determined to minimize the aperture-plane phase variance for each science direction $\bm{\theta_k}$ \cite{Correia-14},
\begin{equation}
\bm{E_{\theta_k}}=\underset{\bm{E_{\theta_k}}}{\text{arg}\text{min}}\langle ||\bm{\varphi_k}-\bm{\hat{\varphi}_k}||^2\rangle,
\end{equation}
where $\bm{E_{\theta_k}}$ is the tomographic reconstructor for the direction $\bm{\theta_k}$, $\bm{\varphi_k}$ is the actual aperture-plane phase coming from the direction $\bm{\theta_k}$, $\bm{\hat{\varphi}_k}=\bm{E_{\theta_k}}\bm{s}$ is the aperture-plane phase estimated from slopes measured by SH-WFSs $\bm{s}$, and $\langle \rangle$ indicates ensemble average over time. With \eref{eq2A-1}, this minimization can be equivalent to the minimization of the variance of the total residual phase distortion in all $N_l$ layers,
\begin{equation}\label{eq2A-5}
\begin{aligned}
\bm{E_{\theta_k}}&=\bm{P_{\theta_k}}\bm{E_s}\\
&=\bm{P_{\theta_k}}\underset{\bm{E_s}}{\text{arg}\text{min}}\langle ||\bm{\phi}-\bm{\hat{\phi}}||^2\rangle,
\end{aligned}
\end{equation}
where $\bm{E_s}$ is the reconstructor providing the phase distortion in each layer.\par
The slope $\bm{s_{\theta_j}}(t)$ measured by the $j$-th SH-WFS aiming the direction $\bm{\theta_j}$ at time $t$ is defined as 
\begin{equation}\label{eq2A-2}
\bm{s_{\theta_j}}(t)=\bm{\Gamma_{\theta_j}}\bm{P_{\theta_j}}\bm{\phi}(t)+\bm{\eta_{\theta_j}}(t),
\end{equation}
where $\bm{\Gamma_{\theta_j}}$ is a discrete phase-to-slope operator which converts phases into slopes and $\bm{\eta_{\theta_j}}(t)$ is a column vector of the noise in measurements of $j$-th WFS. Concatenating \eref{eq2A-2} of $N_{gs}$ WFSs, we can obtain an equation connecting the phase distortion in $N_l$ atmospheric layers and the slope provided by $N_{gs}$ WFSs as follows.
\begin{equation}\label{eq2A-3}
\bm{s}(t)=\bm{\Gamma}\bm{P_{gs}}\bm{\phi}(t)+\bm{\eta}(t),
\end{equation}
where $\bm{s}(t)=[\bm{s_{\theta_1}}^T \cdots \bm{s_{\theta_{N_{gs}}}}^T]^T$, $\bm{\Gamma}$ is  a block diagonal matrix as $\bm{\Gamma}=\textbf{diag}[\bm{\Gamma_{\theta_1}} ,\cdots,\bm{\Gamma_{\theta_{N_{gs}}}}]$,  $\bm{P_{gs}}$ is a GS ray-tracing matrix which is concatenation of $\bm{P_{\theta_j}}$ for $1\leq j\leq N_{gs}$, and $\bm{\eta}(t)$ is a column vector including measurement noises from all WFSs.\par
As shown in \cite{Ellerbroek-02}, the reconstructor can be obtained by minimising the partial derivative of \eref{eq2A-5} with respect to $\bm{E_s}$:
\begin{equation}\label{eq2A-4}
\begin{aligned}
\bm{\hat{\varphi}_{\theta_k}}(t)&=\bm{P_{\theta_k}}\bm{E_s}\bm{s}(t)\\
&=\bm{P_{\theta_k}}\left(\bm{P_{gs}}^T\bm{\Gamma}^T\bm{\Sigma_\eta}^{-1}\bm{\Gamma}\bm{P_{gs}}+\bm{L}^T\bm{L}\right)^{-1}
\bm{P_{gs}}^T\bm{\Gamma}\bm{\Sigma_\eta}^{-1}\bm{s}(t),
\end{aligned}
\end{equation}
where $\bm{\Sigma_\eta}$ is a noise covariance matrix which is a block diagonal matrix as $\bm{\Sigma_\eta}=[\sigma^2_{\eta,\theta_1}\bm{I}, \cdots, \sigma^2_{\eta,\theta_{N_{gs}}}\bm{I}]$ under the assumption that the measurement noise from $j$-th WFS is a zero-mean Gaussian noise with a variance of $\sigma^2_{\eta,\theta_j}$ and that noises of all WFS subapertures are independent of each other. The term $\bm{L}^T\bm{L}$ is an approximation of the phase inverse covariance matrix, $\bm{\Sigma_\phi}^{-1}$, presented by \cite{Ellerbroek-02} and the scaling of $\bm{L}$ is given in \cite{Gilles-13}. While this approximated regularization matrix is sparse and incorporated into efficient sparse-matrix techniques, the approximated matrix does not regularize tip-tilt modes and other modes that are curvature-free \cite{Lee-07}. The unregularized modes, mainly tip-tilt modes, may affect the tomographic error especially in the case with LGSs, where tip-tilt modes can not be measured by the LGS.\par
In the classical tomographic reconstruction, the phase distortion on the turbulence layers is estimated from the measured slopes at one time-step. So, $\bm{E_s}$ is referred to as \textit{single time-step tomographic reconstruction} in the remainder of this paper. 
%
\subsection{Multi Time-Step  Tomographic Reconstruction}
Our idea is to reduce the geometric tomographic error by increasing information of the atmospheric turbulence using the measurements at both the previous and the current time-steps simultaneously. The frozen flow assumption allows us to consider the evolution of the atmospheric turbulence as the movements of the turbulence layers due to the winds. Under this assumption, it can be considered that the areas measured by WFSs at a previous time-step shift due to the winds with time, which is indicated by dashed circles in \fref{fig:MOAO}. Thanks to the spatial displacements due to the winds, the areas corresponding to the measurements at the previous time-step cover the \textit{uncovered} and the \textit{unoverlapped areas} at the current time-step.\par
If an atmospheric turbulence layer moves with a constant velocity $\bm{v}=(v_x, v_y)$, the movement of the phase distortion pattern with a position $x$ at $\Delta t$ previous time-step is written by 
\begin{equation}\label{eq2B-1}
\phi(\bm{x},t-\Delta t)=\phi(\bm{x}+\bm{v}\Delta t,t),
\end{equation}
where $\Delta t$ is the time difference between the previous and the current time-step. Considering this movement, we define the model connecting measurements at the previous time-step ($t-\Delta t$) with the phase distortion due to the atmospheric turbulence at current time-step $t$, 
\begin{equation}\label{eq2B-2}
\bm{s}(t-\Delta t)=\bm{\Gamma}\bm{P_{gs}^{\Delta t}}\bm{\phi}(t)+\bm{\eta}(t-\Delta t),
\end{equation}
where, $\bm{P_{gs}^{\Delta t}}$ is a GS ray-tracing matrix considering the movement of the phase distortion pattern within the GS footprints during the duration time $\Delta t$. The reconstructor, which reconstructs the phase distortion from the measurements at both the current and the previous time-steps, is given as the minimum variance solution of the concatenation of \eref{eq2A-3} with \eref{eq2B-2}. The solution is given as in \eref{eq2A-4}
\begin{equation}\label{eq2B-3}
\begin{aligned}
\bm{\hat{\varphi}_k}(t)&=\bm{P_{\theta_k}}\bm{E_m}\bm{\bar{s}}(t)\\
&=\bm{P_{\theta_k}}\left(\bm{\bar{P}_{gs}}^T\bm{\bar{\Gamma}}^T\bm{\bar{\Sigma}_\eta}^{-1}\bm{\bar{\Gamma}}\bm{\bar{P}_{gs}}
+\bm{L}^T\bm{L}\right)^{-1}\bm{\bar{P}_{gs}}^T\bm{\bar{\Gamma}}\bm{\bar{\Sigma}_\eta}^{-1}\bm{\bar{s}}(t),
\end{aligned}
\end{equation}
where
\begin{equation}
\begin{aligned}
\bm{\bar{s}}=
\begin{bmatrix}
\bm{s}(t)\\
\bm{s}(t-\Delta t)
\end{bmatrix},\ \ \ \ 
\bm{\bar{P}_{gs}}=
\begin{bmatrix}
\bm{P_{gs}}\\
\bm{P_{gs}^{\Delta t}}
\end{bmatrix},\\
\bm{\bar{\Gamma}}=
\begin{bmatrix}
\bm{\Gamma} && \bm{0}\\
\bm{0} && \bm{\Gamma}
\end{bmatrix},\ \ \ \ 
\bm{\bar{\Sigma}_\eta}=
\begin{bmatrix}
\bm{\Sigma_\eta} && \bm{0}\\
\bm{0} && \bm{\Sigma_\eta}
\end{bmatrix}.
\end{aligned}
\end{equation}
Although the multi time-step reconstructor in \eref{eq2B-3} uses measurements only from two time-steps, it is easy to expand the reconstructor to use multi time-step measurements more than two time-steps.\par
The time difference between the current and the previous time steps, $\Delta t$, is an important parameter for the multi time-step reconstructor. Although we make the frozen flow assumption, in reality, the time scale in which the frozen flow assumption is valid is limited and the turbulence changes their structure with time, that is, the turbulence is \textit{boiling} with time. Therefore, $\Delta t$ should be within the time scale in which the frozen flow assumption is valid. Otherwise, the multi time-step reconstructor does not work. Sch{\"o}ck et al. \cite{Schock-00} found the intensity of the cumulative auto-correlation of SH-WFS measurements, including information from multiple turbulence layers, decreases to 90 \% of its initial value in $\sim$25 ms, and to 50 \% in the range of 50 to 100 ms. In this paper, we use this decay ratio as an indicator whether the frozen flow assumption is valid or not.\par
With respect to the computational complexity, the multi time-step reconstructor roughly doubles the size of the matrices. However, all matrices in the multi time-step reconstructor are sparse, and, therefore, a conjugate-gradient iterative scheme with preconditioners \cite{Gilles-02,Yang-06}, warm-started wavefront reconstruction\cite{Lessard-08}, and general purpose computing on GPU (GPGPU) can be used for accelerating the computation of the multi time-step reconstruction.
%
\subsection{Analytical Evaluation of Tomographic Error}
The tip-tilt removed tomographic error in the direction $\bm{\theta_k}$ caused by a reconstructor $\bm{E}$ is defined as (see Appendix A)
\begin{align}\label{eq2C-1}
\sigma^2_\text{tomo}(\bm{\theta_k})&=\langle ||\bm{\varphi_k}-\bm{\hat{\varphi}_k}||^2\rangle/n_\text{node}\notag\\
&=\textbf{Tr}\left[\bm{T}\bm{P_{\theta_k}}(\bm{I}-\bm{E}\bm{\Gamma}\bm{P_\text{gs}})\bm{\Sigma_\phi}(\bm{I}-\bm{E}\bm{\Gamma}\bm{P_\text{gs}})^T\bm{P_{\theta_k}}^T\bm{T}^T\right]/n_\text{node}\notag\\
&+\textbf{Tr}\left[\bm{T}\bm{P_{\theta_k}}\bm{E}\bm{\Sigma_\eta}\bm{E}^T\bm{P_{\theta_k}}^T\bm{T}^T\right]/n_\text{node}\notag\\
&=\sigma_\text{geo}^2+\sigma_\text{noise}^2
\end{align}
where $n_\text{node}$ is the number of valid nodes on the aperture-plane, $\bm{I}$ is an identity matrix, and $\bm{T}$ is a matrix removing a piston and tip-tilt modes. The first term $\sigma_\text{geo}$ represents an error depending on the geometry of GSs and the atmospheric turbulence. The influences from the \textit{uncovered} and \textit{unoverlapped area} are included in the first term. The second term $\sigma_\text{noise}$ represents the propagation of the measurement noise through the tomographic reconstruction.\par
\eref{eq2C-1} can be used for evaluating the tomographic error analytically. Here, we consider a very simple model of a MOAO system with a 30 m circular aperture, 3 NGSs on a ring with a radius from 20 arcseconds up to 200 arcseconds from the center of the scientific FoV, Fried parameter $r_0$ is 0.156 m, an outer scale $L_0$ is 30 m, and 3 atmospheric turbulence layers at 0, 5, and 10 km. The $C_N^2$ fractions are [50 \%, 25 \%, 25 \%], the wind speeds are [5 \ms, 10 \ms, 20 \ms], and the wind directions are [90$^\circ$, 0$^\circ$, 0$^\circ$]. The size of the SH-WFS subaperture is 1 m on the aperture-plane. We assume that all subapertures have the same measurement noise $\sigma_\eta$, therefore $\Sigma_\eta=\sigma_\eta^2\bm{I}$. The spatial sampling of discrete grids on the turbulence layers and the aperture-plane is 1 m, which is same as the subaperture size.\par
The boiling of the turbulence is not taken into account in this analysis, but $\Delta t$ should be determined with considering the time scale in which the frozen flow holds. Since more than half of the turbulence still correlates at 50 ms as mentioned in the 2.B, the frozen flow assumption may mostly or partially hold within this frame. We thus set $\Delta t=50$\,ms and will discuss the time scale in which the frozen flow is valid in more detail in Section 3.\par
\begin{figure}
\centering
\includegraphics[scale=1]{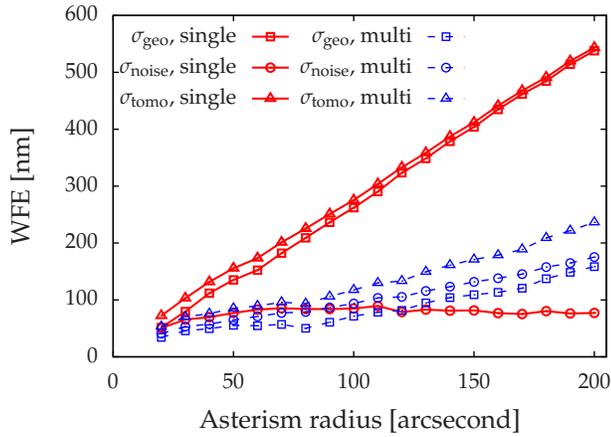}
\caption{Tomographic error $\sigma_\text{tomo}$ (triangles), geometric error $\sigma_\text{geo}$ (squares), and propagation of measurement noise $\sigma_\text{noise}$ (circles) computed by \eref{eq2C-1} with the single time-step reconstructor (the symbols with the red solid line) and the multi time-step reconstructor (the symbols with the blue dashed line). The fitting error is not included in the results.}
\label{fig:static}
\end{figure}

\fref{fig:static} shows $\sigma_\text{tomo}$, $\sigma_\text{geo}$, and $\sigma_\text{noise}$ in the central direction computed by \eref{eq2C-1} with the single (the red solid lines in the figure) and multi (the blue dashed lines) time-step reconstructors with different asterism radii. For the single time-step reconstructor, the geometric error $\sigma_\text{geo}$ increases largely with the radius of the asterism due to the increase in the \textit{unoverlapped area}, and dominates the total tomographic error $\sigma_\text{tomo}$. On the other hand, in the case of the multi time-step reconstructor, both of the geometric error $\sigma_\text{geo}$ and the noise propagation $\sigma_\text{noise}$ increase with the asterism radius, but the errors are much smaller than $\sigma_\text{geo}$ of the single time-step reconstructor. Consequently, the multi time-step reconstructor reduces $\sigma_\text{tomo}$ especially at larger asterism in comparison with the single time-step reconstructor. This result shows that the multi time-step reconstructor has a potential to reduce the tomographic error by using the measurement at previous time steps and to enable the expansion of the size of scientific FoV of WFAO systems without reducing the correction performance. It is noted that this analytical tomographic error is based on the simple model and does not include the fitting error, the temporal lag-error, uncertainty of the atmospheric turbulence model, and uncertainties of the wind speeds and directions.\par
\begin{figure}
\centering
\includegraphics[scale=1]{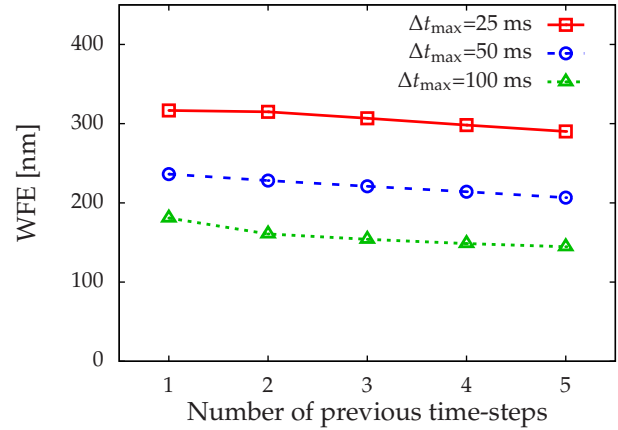}
\caption{Tomographic error $\sigma_\text{tomo}$ with the NGS asterism of 200 arcsecond radius as a function of the number of previous time steps, $N_{\Delta t}$. The variation of the line colors and types indicate the difference of the maximum $\Delta t$, which is denoted as $\Delta t_\text{max}$. Each time difference is the integral multiple of $\Delta t_\text{max}/N_{\Delta t}$. For example, if the number of the previous time-steps is 4 and $\Delta t_\text{max}$ is 100 ms, the time differences used in the multi time-step reconstruction are 25 ms, 50 ms, 75 ms, and 100 ms.}
\label{fig:static_multi}
\end{figure}

\fref{fig:static_multi} shows the dependency of the tomographic error $\sigma_\text{tomo}$ on the number of the previous time-steps, $N_{\Delta t}$. Different lines show the results with different maximum time differences, $\Delta t_\text{max}$, used in the multi time-step reconstruction. The temporal sampling of $\Delta t$ is determined by $\Delta t_\text{max}/N_{\Delta t}$. Namely, if $N_{\Delta t}$ is 4 and $\Delta t_\text{max}$ is 100 ms, the time differences used in the multi time-step reconstruction are 25\,ms, 50\,ms, 75\,ms, and 100\,ms. The tomographic error decreases with the number of the previous time-steps for the same $\Delta t_\text{max}$. Also, the tomographic error is reduced with longer $\Delta t_\text{max}$. The gain with using longer $\Delta t_\text{max}$ is larger than the gain with increasing the number of the time-steps, if the frozen flow assumption holds. In other words, the time scale in which the frozen flow assumption is valid has the most significant impact on the multi time-step reconstruction. We can reduce the tomographic error by increasing the number of the previous time-steps, as much as  the computational complexity of the multi time-steps is acceptable.
%
\section{End-to-End Simulation}
%
\subsection{Setting}
\begin{figure}
\centering
\includegraphics[scale=1]{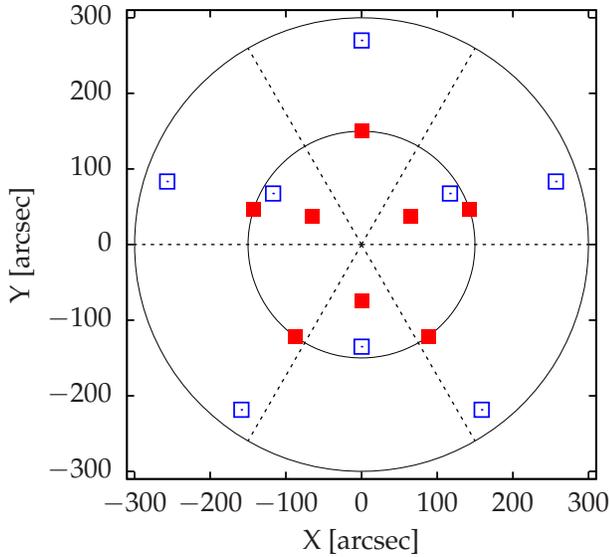}
\caption{LGS asterisms used in the numerical simulation. A narrow asterism is indicated with the red filled squares. The blue open squares show a wide asterism. The black solid lines show the 2.5 arcminutes and 5 arcminutes radius from the center of a scientific FoV. The dashed lines are directions where the performance is evaluated.}
\label{fig:simu_asterism}
\end{figure}

Parameters used in the E2E simulation are listed in \tref{tab:setting}. We assume a MOAO system with a scientific FoV with a 10 arcminute diameter on a 30 m circular-aperture telescope with 8 sodium LGSs at 90 km. Two asterisms are used in the simulation, which are shown in \fref{fig:simu_asterism}. One is a narrow asterism, indicated as open squares in \fref{fig:simu_asterism}, and the other is a wide asterism, indicated as filled squares. Low-order modes of the phase distortion, which are tip, tilt and focus, can not be measured by an LGS due to uncertainty in LGS position on sky. Therefore, an NGS is required in a scientific FoV to measure these low-order modes when we use an LGS. In this simulation, however, we assume that we can measure the low-order modes of the phase distortion from the LGSs for simplicity. The cone effect and the spot elongation on SH-WFSs are considered in the simulation. The sodium-layer profile is approximated to the lidar measurements used in \cite{Gilles-06} as
\begin{equation}
\text{Na}(h)=\left\{
\begin{matrix}
\exp\left\{-\frac{(h-h_\text{LGS})^2}{2\sigma_\text{Na}^2}\right\}&(\text{for\ \ }|h-h_\text{LGS}|<\sigma_\text{Na})\\
0&(\text{for\ \ }|h-h_\text{LGS}|>\sigma_\text{Na})
\end{matrix}
\right.
\end{equation}
where $h_\text{LGS}=88$ km is the altitude of a sodium layer above the telescope, and $\sigma_\text{Na}$=5 km is the half width of the sodium-layer. We use SH-WFSs with 60$\times$60 subapertures and DMs for science targets with 60$\times$60 elements.\par
We assume the top of Maunakea as the observation site in the simulation and use a seven-layer model used in \cite{Andersen-12} for an atmospheric turbulence profile. This model is created based on image-quality measurements from the Subaru Observatory \cite{Miyashita-04}, combined with differential image motion monitor (DIMM) and multiple aperture scintillation sensor (MASS) measurements by the Thirty Meter Telescope Project at Maunakea \cite{Els-09}. The model includes the additional dome seeing in the Subaru telescope, thus the ground layer has a strong turbulence power, which is 60 \% of the total turbulence power. Fried parameter $r_0$ is in the 50th percentile of the seeing measurements, $r_0=0.156$ m, and an outer scale is assumed to be 30 m. The $C_N^2$ values at each altitude are summarized in \tref{tab:atmos_simu}. We assume a gaussian model for the wind speeds on each layer based on \cite{Hardy-98}. The assumed wind speeds and directions are also listed in \tref{tab:atmos_simu}. The performance is evaluated at wavelength of 1650 nm (H-band). The turbulence simulated in the simulation follows perfectly the frozen flow assumption. 
%
\subsection{Simulation Results}
\begin{figure*}
\centering
\includegraphics[scale=1]{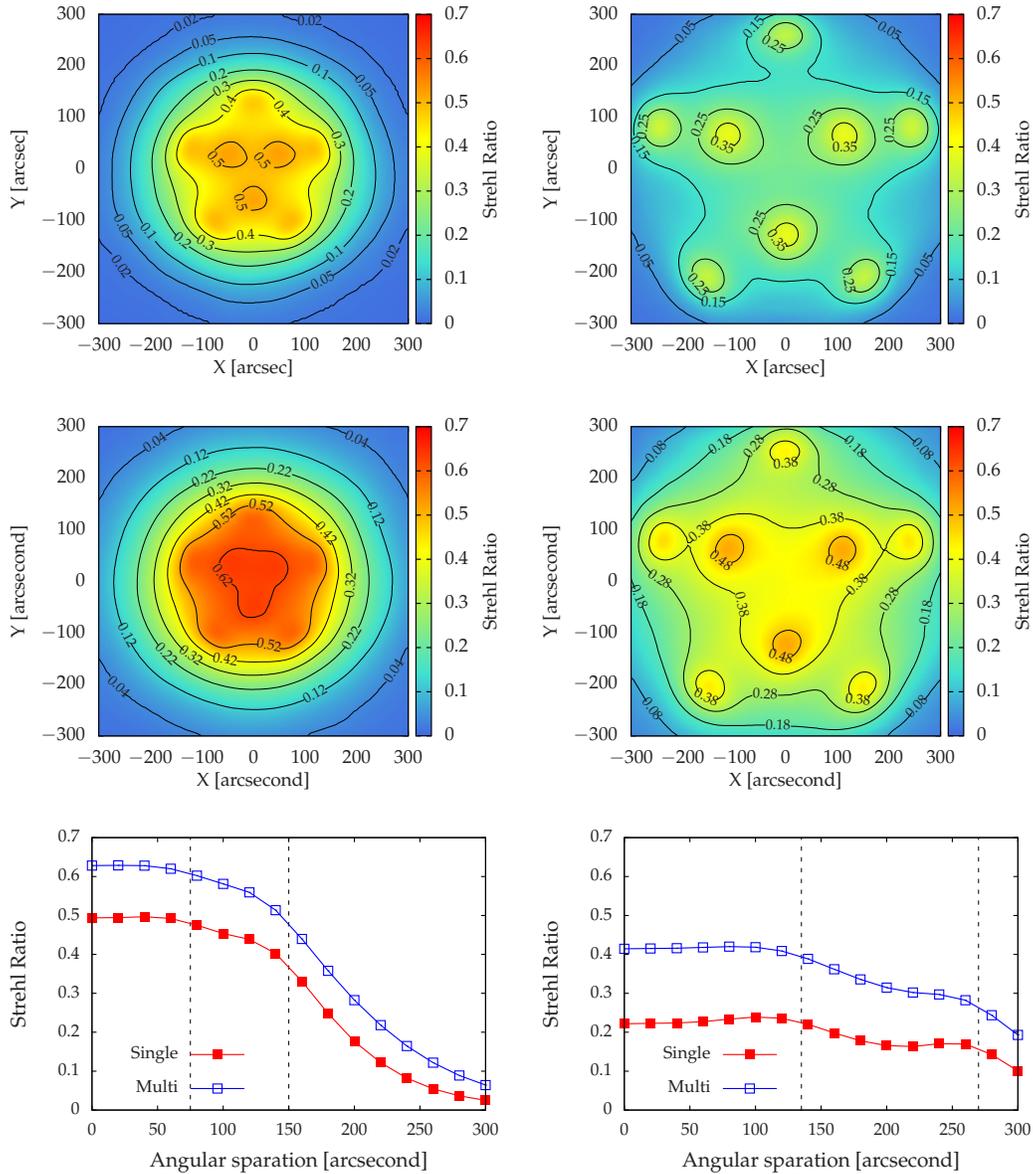}
\caption{Top and middle: Simulated SR maps within the 10 arcminutes scientific FoV computed by using the single time-step reconstructor (top) or the multi time-step reconstructor with $\Delta t$=50 ms (middle). The right panels are the results with the narrow LGS asterism, and the left panels are computed with the wide LGS asterism. The multi time-step reconstructor improves SR values over the scientific FoV for both of the asterisms. Bottom: SR profiles as a function of an angular separation from  the center direction of the scientific FoV. The profiles are averaged over directions shown as the dashed lines in \fref{fig:simu_asterism}. The red filled squares show the averaged SR profile with the single time-step reconstructor, and the blue open squares is the result by the multi time-step reconstructor with $\Delta t$=50 ms. The dashed black lines show the radii of LGS positions. The SR values shown in this figure are evaluated with including the tomographic error, the fitting error, and the temporal lag error. The turbulence used in the simulation follows perfectly the frozen flow.}
\label{fig:SRmap_profile}
\end{figure*}

First, we simulated the optimal performance of the single and the multi time-step reconstructors for both the narrow and the wide LGS asterisms with an assumption that the turbulence profile, wind speeds and directions are known \textit{a priori}. The results are illustrated in \fref{fig:SRmap_profile}. SR values are measured from simulated point spread function (PSF) images by comparing the peak intensities of the PSF images to the peak intensity of a diffraction-limited PSF image created by the simulation. All PSF images from the simulation are normalized by the total intensity within a 1 arcsecond box. The top and the bottom panels in \fref{fig:SRmap_profile} show the simulated maps of the Strehl ratio (SR) across the scientific FoV of 10 arcminutes diameter with the single (top) and the multi time-step (bottom) reconstructors. We use $\Delta t$ of 50 ms for the multi time-step reconstructor as in Section 2. The bottom panels in \fref{fig:SRmap_profile} show SR profiles as a function of an angular distance from the center of the scientific FoV. The profiles are computed by averaging over six directions shown as dashed lines in \fref{fig:simu_asterism}.\par
With the narrow asterism, the single time-step reconstructor, which is indicated as red filled squares in the left panels of \fref{fig:SRmap_profile}, achieves an average SR of $\sim$0.5 within the inner LGS radius at 75 arcseconds from the center. The average SR value decreases slowly with an angular separation from 75 arcseconds to 150 arcseconds. In the region outside 150 arcseconds radius, the average SR values decrease steeply because there is no LGS. The multi time-step reconstructor with $\Delta t=$50 ms, represented as the blue open squares, can increase the SR value over the scientific FoV, The improvement factor of the average SR value is 1.25 at 150 arcseconds, even though there is no \textit{uncovered area} at the inner region below 150 arcseconds. This suggests that the single time-step reconstructor is affected by the degeneracy due to the \textit{unoverlapped area} even with the narrow asterism.\par
In the case of the wide asterism, illustrated in the right panels of \fref{fig:SRmap_profile}, the SR value in the region outside 200 arcseconds radius is better than the value of the narrow asterism because the wider area is covered by the LGSs. In the contrast, the larger separation between LGSs causes a larger \textit{unoverlapped area} at high altitudes. Hence, the SR value of the inside area becomes less than that of the narrow asterism, and there are valley areas between LGSs on the SR map (the left top panel of \fref{fig:SRmap_profile}). From the bottom right panel of \fref{fig:SRmap_profile}, the maximum average SR value is 0.24 with the single time-step reconstructor. This is affected strongly by \textit{unoverlapped area}. Using the multi time-step reconstructor with $\Delta t=50$ ms,  the average SR value is doubled over the scientific FoV for the wide asterism. At 300 arcseconds from the center, the SR value from the multi time-step reconstructor with the wide asterism is 3 times larger than the result of the narrow asterism.\par
\begin{figure}
\centering
\includegraphics[scale=1]{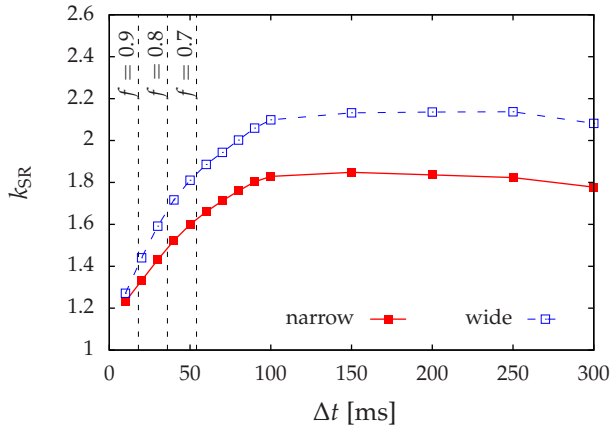}
\caption{Dependence of the multi time-step reconstructor on the $\Delta t$ is presented in \fref{fig:SRdiff}. The vertical axis in \fref{fig:SRdiff} is the SR improvement ratio, $k_\text{SR}$, which represents the ratio of the SR value achieved by the multi time-step reconstructor to the SR value from the single time-step reconstructor and is averaged over the angular separation from the center of the FoV. This SR improvement ratio includes the tomographic error, the fitting error, and the temporal lag error. The result with the narrow asterism is indicated as the red filled squares, and the blue open squares shows the result with the wide asterism. The vertical dashed lines indicate the time duration that the decay ratio $f$ of the temporal correlation of SH-WFS measurements is 0.9, 0.8, and 0.7 when the wind speed is 33 \ms, which is maximum value in the wind model, estimated from \eref{eq3B-1}.}
\label{fig:SRdiff}
\end{figure}

The performance of the multi time-step reconstructor depends on the spatial displacement between the areas measured by the current and the previous time-step at each altitude, which we refer as $d(h)$ in the paper. This displacement $d(h)$ corresponds to the movement of the atmospheric turbulence layer at an altitude $h$ during $\Delta t$, $d(h)=v(h)\Delta t$ with $v(h)$ the wind speed at altitude $h$. The dependence of the multi time-step reconstructor on the $\Delta t$ is presented in \fref{fig:SRdiff}. The vertical axis in \fref{fig:SRdiff} is the SR improvement ratio, $k_\text{SR}$, which represents the ratio of the SR value achieved by the multi time-step reconstructor to the SR value from the single time-step reconstructor and is averaged over the angular separation from the center of the scientific FoV. The multi time-step reconstructor achieves larger $k_\text{SR}$ for the wide asterism than the narrow one. This suggests the wide asterism is affected more by the \textit{unoverlapped area} than the narrow asterism. However, the trend of the dependence of $k_\text{SR}$ on $\Delta t$ looks similar for both asterisms. This suggests that the dependence on $\Delta t$ doesn't depend the asterism of GSs. The multi time-step reconstructor improves the SR even with small $\Delta t$. The improvement ratio is 1.23 for $\Delta t$=20 ms for the narrow asterism. The improvement ratio $k_\text{SR}$ is maximized at $\Delta t\sim$100 ms, where $k_\text{SR}\sim$1.8 and $\sim$2.1 for the narrow and the wide asterism, respectively. The displacements $d$ with $\Delta t$=100 ms are 3.3 m and 0.7 m for the fastest and slowest layers in the model. While, as $\Delta t$ is larger than 100 ms, $k_\text{SR}$ has almost no or very week dependency on $\Delta t$. This suggests that only small displacements between the areas measured at the current and the previous time-step is enough to solve the degeneracy due to the \textit{unoverlapped area}.\par
\begin{figure}
\centering
\includegraphics[scale=1]{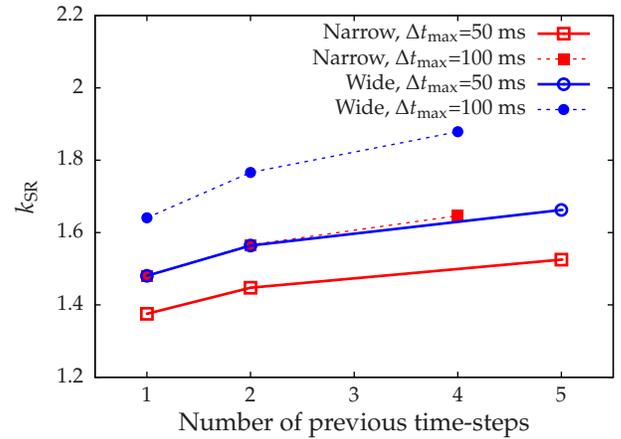}
\caption{The improvement ratio of SR $k_\text{SR}$ as a function of the number of previous time steps, $N_{\Delta t}$. The red circles show the result with the narrow asterism and the blue squares show the result with the wide asterism. The maximum time differences, $\Delta t_\text{max}$, are set to 50 ms (the open symbols) and 100 ms (the filled symbols).}
\label{fig:SRmulti}
\end{figure}

\fref{fig:SRmulti} shows the dependency of the improvement ratio $k_\text{SR}$ on the number of the previous time-steps, $N_{\Delta t}$. The improvement ratio increases with $N_{\Delta t}$, which is the same trend shown in the analytical evaluation (\fref{fig:static_multi}). Increasing the number of the previous time-steps provides the improvement in SR, if the computational complexity is acceptable.\par
The results shown above assume that the atmospheric turbulence follows perfectly the frozen flow assumption. As mentioned before, however, the time scale of the frozen flow assumption is valid only within a short time. Beyond this range the phase distortion pattern of the atmospheric turbulence varies. If the time scale of the frozen flow is smaller than $\Delta t$ for the multi time-step reconstructor, the measurement at the previous time step can no longer be used for it is uncorrelated with the current measurement. Guesalaga et al. \cite{Guesalaga-14} investigated the time evolution of the atmospheric turbulence by using spatio-temporal cross-correlations of the measurements from multiple SH-WFSs installed in the Gemini South Multi-Conjugate Adaptive Optics System (GeMS). They found that the decay ratio of the correlation peak intensity for an individual layer, $f$, decreases with the time delay for the correlation, $\Delta t$, and depends on the distance travelled by the layer, $d$. In their paper, the decay ratio is approximated as a function of $\Delta t$ and $v$.
\begin{equation}\label{eq3B-1}
f=(-0.157v-0.365)\Delta t+1
\end{equation}
By using this equation, we estimate $\Delta t$ to restrict the decay ratio $f$ to 0.9, 0.8, and 0.7 assuming the fastest wind speed in the model, $v=33$ \ms, which is plotted in \fref{fig:SRdiff} as vertical dashed lines. In order to keep the decay ratio greater than 0.7 for the fastest layer, $\Delta t$ should be smaller than 54 ms. For the slower layer, required $\Delta t$ to make the decay ratio 0.7 gets larger. In this paper, we assume that the frozen flow assumption is valid when the decay ratio is greater than 0.7 and set $\Delta t$ to keep the decay ratio greater than 0.7. However, it is still unclear how the boiling of the turbulence affects on the performance of the multi time-step reconstruction, which is not modeled in the simulation. In order to understand the limits of the frozen flow assumption and evaluate the effect of the temporal de-coherence of the turbulence, on-sky experiments are necessary.\par
\begin{figure}
\centering
\includegraphics[scale=1]{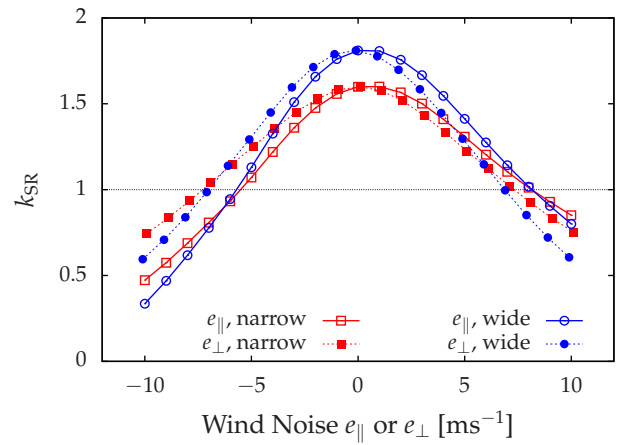}
\caption{SR improvement ratio, $k_\text{SR}$, achieved by multi time-step tomographic reconstruction with different \textit{wind speed errors} $e_\parallel$ (the filled symbols) or \textit{wind direction errors} $e_\perp$ (the open symbols). The results with the narrow asterism are represented as the red squares and the results with the wide asterism are the blue circles. The SR ratio less than 1 means that the performance of the multi time-step reconstructor is poorer than the performance achieved by the single reconstructor due to the wind estimation error.}
\label{fig:SRerror}
\end{figure}

At the end of this section, we discuss the effect of uncertainties of wind speed and direction. The errors in wind speeds and directions result in an error of the spatial displacement between the areas measured by the current and the previous time-step. Here, we represent the wind error as $\bm{e}=(e_\parallel, e_\perp)$. The first one, $e_\parallel$, is an error parallel to the wind direction and referred as \textit{wind speed error}. The second, $e_\perp$, is an error perpendicular to the wind direction and referred as \textit{wind direction error}. \fref{fig:SRerror} shows the improvement ratio $k_\text{SR}$ achieved by the multi time-step reconstructor compared to the single time-step reconstructor with different errors of wind speed and direction, for both the narrow (red symbols) and wide (blue symbols) asterisms. We assume the same wind error for all atmospheric layers and $\Delta t$=50 ms for the multi time-step reconstructor. The curve of $k_\text{SR}$ with the \textit{wind direction error} $e_\perp$ is symmetric with respect to the point $e_\perp=0$. The multi time-step reconstructor has the advantage compared to the singe time-step reconstructor, $k_\text{SR}>0$, as $-$7 \ms\ $\leq e_\perp\leq$ 7 \ms\ for the condition we use for both asterisms. This corresponding spatial displacement error is less than 0.35 m and the error in angle of wind direction is less than 12 degrees for the fastest layer (33 \ms) and 45 degrees for the slowest layer (7 \ms) wind. For the \textit{wind speed error} $e_\parallel$, the multi time-step reconstructor is affected more by negative $e_\parallel$, which is an error in the opposite direction to the wind, than positive $e_\parallel$. This is because if $e_\parallel$ is negative, the covered areas are smaller than if $e_\parallel$ is positive. The allowable range for $e_\parallel$ is $-$5.5 \ms\ $\leq e_\parallel\leq$ 8 \ms\ for the wide asterism and $-$5 \ms\ $\leq e_\perp\leq$ 7 \ms\ for the narrow asterism. The dependence of the multi time-step reconstructor on the wind errors depends on the turbulence power, wind speed, direction and $\Delta t$. If the turbulence powers are stronger, the influence of the uncertainty of wind speed and direction is larger. Using larger $\Delta t$ also makes the tomographic error due to the wind uncertainty larger because the spatial displacement error increases.
%
\section{Estimation of Wind Speed and Direction}
As mentioned before, the altitude and power of the atmospheric turbulence layers (the turbulence profile) and the wind speeds and directions of multiple layers (the wind profile) are essential prior information for the multi time-step reconstruction. The turbulence profile can be estimated from the spatial cross-correlation of SH-WFS measurements by the SLOpe Detection And Ranging (SLODAR), which is well studied method and already tested on sky \cite{Wilson-02,Butterley-06,Cortes-12}. We estimate the turbulence profile by the SLODAR proposed in \cite{Butterley-06}.\par
The wind profile directions can be estimated by the spatio-temporal cross-correlation of the SH-WFS measurements \cite{Guesalaga-14}. The multiple peaks on the spatio-temporal cross-correlation map, which correspond to the multiple turbulence layers at different altitudes, move with the time delay for the spatio-temporal cross-correlation, depending on the wind speeds and directions. By tracking this correlation peaks,  we can estimate the direction and speed of the wind of each turbulence layer. However, isolating and tracking the multiple peaks on the spatio-temporal correlation maps will be difficult when the peaks are not isolated, for example, the multiple peaks are overlapped along the tracks. In particular, in order to implement this wind estimation method to the real-time control, the method automatically needs to isolate and detect the multiple peaks.\par
In order to overcome this issue, we propose the tomographic wave-front reconstruction first to isolate the turbulence layer at different altitudes. Then, the spatio-temporal auto-correlation are applied to the isolated turbulence at each altitude to compute the wind speed and direction at its altitude. If the tomographic reconstruction works well, the spatio-temporal auto-correlation map of each turbulence layer has only one peak and we can track the peak more easily than the original spatio-temporal cross-correlation method. Furthermore, once the wind profile is estimated, we can use the multi time-step reconstruction to isolate each turbulence layer more clearly than the single time-step reconstruction and thus improve the accuracy of the wind estimation.\par
\begin{figure*}[p]
\centering
\includegraphics[scale=1]{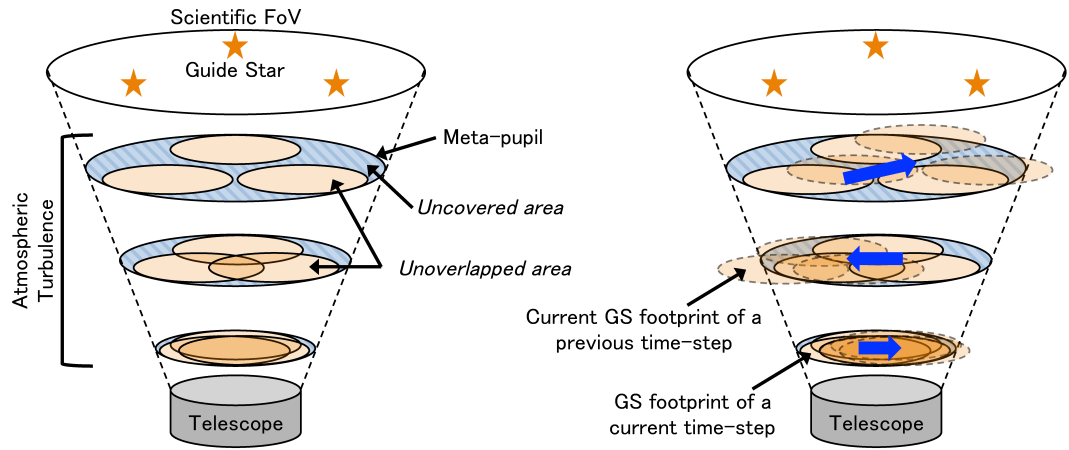}
\caption{Geometric relation between positions of GSs and regions in the atmosphere measured by WFSs. Left: The solid orange circles show the footprint of GS optical paths at the current time-step. There are areas covered by no or only one GS, referred as \textit{uncovered area} or \textit{unoverlapped area}. These areas cause significant tomographic errors. Right: The blue arrows indicate wind direction at each altitude. The GS footprints measured at a previous time-step move due to the wind with time, which are dashed orange circles in the figure. Using the current and previous time-step measurement, which correspond to solid and dashed orange regions, we can reduce the \textit{uncovered} and \textit{unoverlapped areas} and improve the accuracy of a tomographic reconstruction.}
\label{fig:MOAO}
\end{figure*}

After the tomographic reconstruction provides the the phase distortion of each turbulence layer, we extract an aperture-size wavefronts at each altitude in the direction of the center of a GS asterism from the reconstructed phase distortions because the size of the turbulence layers are too large to compute spatio-temporal auto-correlation. In addition, it is better to use only areas reconstructed accurately for estimating wind speeds and directions. Although the GS directions have smaller integrated WFE or better SR than other directions (see the SR maps in \fref{fig:SRmap_profile}), it is possible that most areas at high altitude in the GS directions are covered only by one optical path, as shown in \fref{fig:MOAO}, and are affected by the degeneracy due to the \textit{unoverlapped area}. This is because the reconstructed phase distortion at each altitude in the GS direction can cancel each other out and, as a result, the total WFE in this direction becomes smaller than other directions. Therefore, since the reconstructed phase distortion in the GS directions may not be isolated well, using GS direction for the wind estimation is not optimal. As shown in \fref{fig:MOAO}, the direction of the center of the GS asterism are covered by multiple GS footprints, and the reconstructed phase distortion at each altitude is relatively accurate compared to the other directions.\par
We define a matrix $\bm{P_c}$ as the cropping matrix, which extracts the wavefront at each altitude, not integrated wavefront, in the direction of the center of the GS asterism from the reconstructed phase distortion $\bm{\hat{\phi}}(t)=[\bm{\hat{\phi}_1}^T(t) \cdots \bm{\hat{\phi}_{N_l}}^T(t)]^T$.
\begin{equation}\label{eq4A-1}
\bm{\hat{\phi}_c}(t)=\left[\bm{\hat{\phi}_{c,1}}^T(t)\  \bm{\hat{\phi}_{c,2}}^T(t)\  \cdots\ \bm{\hat{\phi}_{c,N_l}}^T(t)\right]^T=\bm{P_c}\bm{\hat{\phi}}(t).
\end{equation}\par
We perform the temporal auto-correlation of each turbulence layer in the slope space. The power spectral density (PSD) of the phase and slope have dependencies on the spatial frequency, $k$, as $k^{-11/3}$ and $k^{-5/3}$, respectively. Therefore, the temporal correlation of the slope tends to be more heightened at small time-delay and more long-lived at large separation than the phase due to the less steep PSD of the slope of phase. Thus, the peak of the slope correlation can be tracked easily compared with the peak of phase correlation. The conversion to the slope from the phase is performed by using the discrete phase-to-slope operator, $\bm{\Gamma_c}$, as $\bm{\hat{s}_{c,i}}(t)=\bm{\Gamma_c}\bm{T}\bm{\hat{\phi}_{c,i}}(t)$, where $\bm{\hat{s}_{c,i}}(t)$ is a reconstructed slope vector corresponding to a phase distortion of the $i$-th atmospheric turbulence layer in the direction of the center of the GS asterism at time $t$, and $\bm{T}$ is a matrix removing tip-tilt modes, which are affected by vibrations from telescope and instrument.\par
Temporal correlation of the reconstructed slope at the $i$-th turbulence layer can be computed as
\begin{equation}\label{eq4A-3}
C_{i,x}(\delta u, \delta v, \delta t)=\frac{\left\langle\sum_{u,v}\bm{\hat{s}_{c,i,x}^{u,v}}(t)\bm{\hat{s}_{c,i,x}^{u+\delta u,v+\delta v}}(t+\delta t)\right\rangle}{N(\delta u, \delta v)},
\end{equation}
where $u$ and $v$ are the subaperture indices along the $x$ and $y$ directions, respectively, $\bm{\hat{s}_{c,i,x}^{u,v}}(t)$ is an estimated $x$-direction slope vector in the subaperture $(u,v)$ at time $t$, $\delta u$ and $\delta v$ are relative distances between two subapertures, $\delta t$ is the time delay for the temporal correlation. The summation, $\sum_{u,v}$, indicates the sum of all valid subaperture pairs with relative separation ($\delta u$,$\delta v$), $N(\delta u, \delta v)$ is the number of the pair with the separation ($\delta u$,$\delta v$), and $\langle\rangle$ represents average over time. Computing \eref{eq4A-3} for all possible pairs of $\Delta u$ and $\Delta v$, we can obtain a correlation map for the reconstructed $x$-direction slopes of the $i$-th turbulence layer. The correlation for the $y$-direction slopes is represented similarly to \eref{eq4A-3}. In order to increase signal to noise, the correlations of the $x$ and $y$-direction slopes are averaged, $\bm{C_i}=(\bm{C_{i,x}}+\bm{C_{i,y}})$. \par
The movement of the correlation peak on the correlation map $\bm{C_i}$ with $\delta t$ corresponds to the wind speed and direction of the $i$-th turbulence layer. We use the centroid algorithm to detect the peak. With a SH-WFS subaperture size is $d_{sub}$ and the peak position is $(\Delta u, \Delta v)$, the wind speed is computed by $v_x=d_{sub}\Delta u/\delta t$ and $v_y=d_{sub}\Delta v/\delta t$. By computing $v_x$ and $v_y$ and averaging each one over different $\delta t$, the wind speed and direction at each altitude can be estimated.
%
\section{LABORATORY TEST WITH RAVEN}
%
\subsection{RAVEN}
RAVEN is a MOAO technical and science demonstrator installed and tested on the Subaru telescope, which has a 8 m primary mirror, at Maunakea \cite{Olivier-14}. RAVEN can apply MOAO correction simultaneously for two science targets by tomographic reconstruction with 3 NGSs and 1 LGS attached to the Subaru telescope \cite{Hayano-10}. The system of RAVEN is summarized in \cite{Olivier-14}.\par
\begin{figure}
\centering
\includegraphics[scale=1]{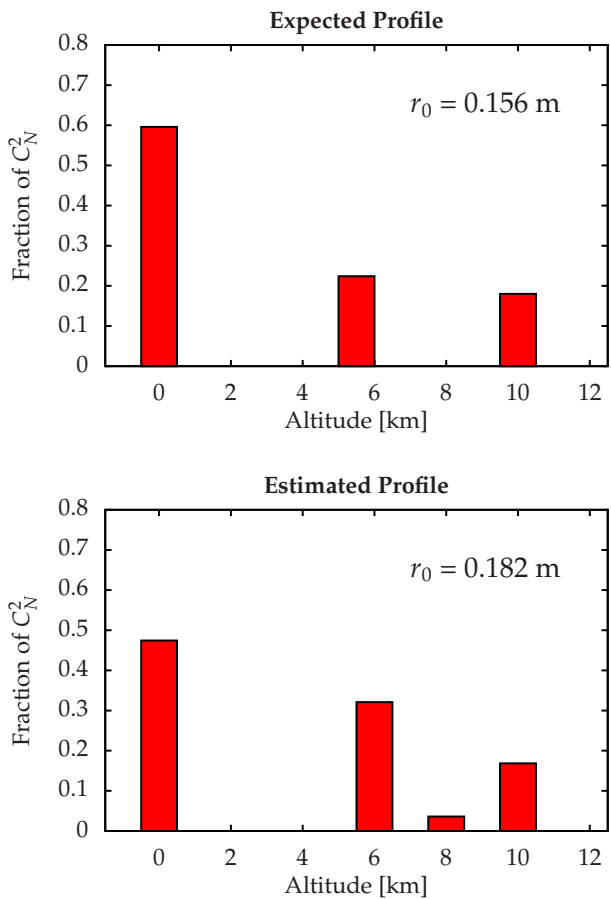}
\caption{Left: The expected turbulence Profile generated by the RAVEN CU. Right: The turbulence profile estimated by the SLODAR in the lab on RAVEN.}
\label{fig:slodar}
\end{figure}

RAVEN has 4 open-loop SH-WFSs (OL-WFS) with 10$\times$10 subapertures in the system, which are for 3 NGSs and 1 LGS. Each science optical path contains an ALPAO DM with 11$\times$11 actuators. RAVEN has a calibration unit (CU) on the optical bench, which simulates multiple GSs and science targets, 3 turbulence layers, and a telescope to calibrate and test the AO system \cite{Lavigne-12}. Two high layers at altitude of 10 km and 5 km are simulated by phase screens, which can be rotated to simulate a movement of the turbulence layers due to the wind. The ground layer at 0 km is generated by a calibration DM with 17$\times$17 actuators. The turbulence parameters simulated by the CU are summarized in the left panel of \fref{fig:slodar} and the top table of \tref{tab:slodar}. It should be noted that the phase distortion patterns generated by the CU are periodic, and the period is 40 s for the 5 km layers and 19 s for the 10 km layer. RAVEN also has a near infrared camera on the optical bench for laboratory experiments.\par
\begin{figure}
\centering
\includegraphics[scale=1]{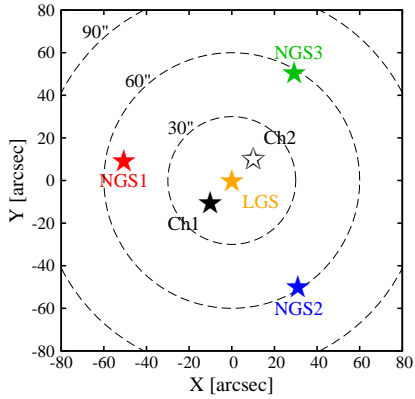}
\caption{Asterism of the GSs and the science targets for the laboratory test with the RAVEN.}
\label{fig:asterism}
\end{figure}

We test the wind estimation method and the multi time-step reconstruction with the CU and the infrared camera installed in RAVEN. An asterism we used is shown in \Fref{fig:asterism}. The NGS are at approximately a 60 arcsecond radius. The brightness of the NGS is set to R$\sim$8 mag, which is very bright and the WFS measurement error is minimal. The frame rate of the control is 250 Hz.
%
\subsection{SLODAR and wind estimation}
\fref{fig:slodar} and \tref{tab:slodar} show comparisons between the expected and estimated turbulence profile. The profile estimated by the SLODAR has three major peaks at the almost same altitudes as that of the expected profile. The estimated power of the ground layer is weaker than the expected value. As a result, the estimated total $r_0$ of 0.182 m is larger than the expected $r_0$ of 0.156 m. Larger estimated $r_0$ than the designed value is already reported in the previous estimation in the lab on RAVEN \cite{Olivier-14}. Since the actuator size of the calibration DM in the CU is roughly 0.47 m, the DM can not reproduce the phase distortion with spatial scales smaller than 0.47 m. We conclude that the actuator size cause the discrepancy and that the estimated value reflects the real turbulence profile on the DM better than the designed value.\par
\begin{figure}
\centering
\includegraphics[scale=1]{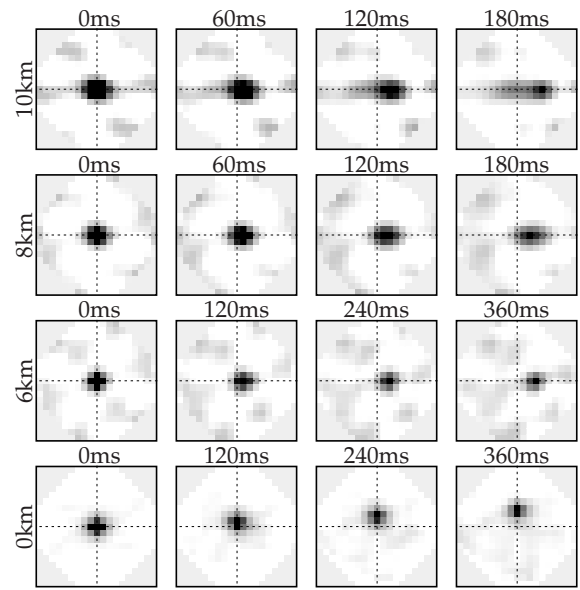}
\caption{Temporal correlation maps of slope at each altitude estimated by tomographic reconstruction. The values at the top of each image indicate the time delay used to drive the temporal correlation.}
\label{fig:corr}
\end{figure}

By using the estimated profile, we evaluate the wind speed and direction at each layer. First, we compute the tomographic reconstruction to reconstruct the phase distortion pattern at each altitude. Then, the temporal correlation map of the estimated slopes are calculated with different delay times, $\delta t$. The temporal correlation maps are illustrated in \fref{fig:corr}. The delay time for the lower layers at 0 km and 6 km are set to larger than the higher layers (\fref{fig:corr}). Generally, the wind speed at low altitude is considered slower than the wind speed at high altitude. In order to get clear movement of the temporal correlation peak, longer delay time is preferred for the low altitude layers. For all layers detected by the SLODAR, the peaks on the temporal correlation maps are detected. The estimated wind speeds and directions, which are summarized in the bottom table of \tref{tab:slodar}, are consistent with the expected value. The error of the wind speed is less than 2 \ms\ for all layers, which is an acceptable wind error computed by the numerical simulation in Section 3. With this wind estimation error, the multi time-step reconstructor will work with almost maximum performance on the future MOAO system (see \fref{fig:SRerror}).\par
\begin{figure*}
\centering
\includegraphics[scale=1]{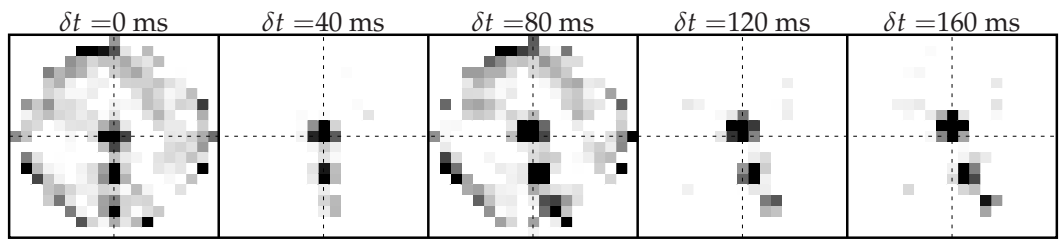}
\caption{Temporal cross-correlation map between the slopes from SH-WFS2 and SH-WFS3. This map is used for the wind estimation proposed in \cite{Guesalaga-14}. There are three peaks, moving with $\delta t$, on the temporal cross-correlation map. The altitudes corresponding to these peaks are 0 km, 6 km, and 11 km, respectively. The wind speeds and directions are estimated by tracking these peaks. The estimated wind speeds in the $x$ and $y$ directions are [-0.1, 4.7] \ms\ at 0 km, [6.6, 0.0] \ms\ at 6 km, and [17.4, 3.0] \ms at 11 km. These values are consistent with the expected values and the values estimated by the wind estimation proposed in this paper.}
\label{fig:slodarwind}
\end{figure*}

We also compute the wind speeds and directions by the method proposed in \cite{Guesalaga-14} as a comparison. The temporal cross-correlation maps between the slopes from SH-WFS2 and SH-WFS3 are shown in \fref{fig:slodarwind}. There are three peaks, moving with $\delta t$, on the temporal cross-correlation maps. The altitudes corresponding to these peaks are 0 km, 6 km, and 11 km, respectively. The wind speeds and directions are estimated by tracking these peaks. The estimated wind speeds in the $x$ and $y$ directions are [-0.1, 4.7] \ms\ at 0 km, [6.6, 0.0] \ms\ at 6 km, and [17.4, 3.0] \ms at 11 km. These values are consistent with the expected values and the values estimated by our wind estimation. In this laboratory test, there is no overlap of the peaks along tracks and thus, tracking the peaks is not difficult. However, if there is more peaks or/and there are the overlaps of the peaks, tracking the peaks becomes more difficult. On the other hand, since the temporal correlation map of the reconstructed slopes (\fref{fig:corr}) provides only one peak in their maps, our method makes tracking the peaks easier. This is important to implement the wind estimation automatically.
%
\subsection{Multi time-step reconstruction}
\begin{figure*}
\centering
\includegraphics[scale=1]{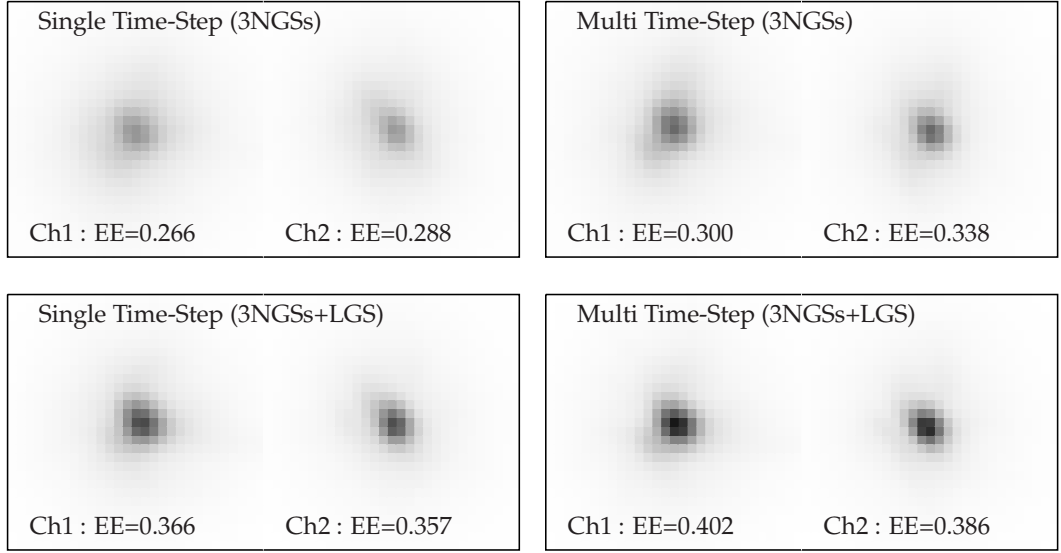}
\caption{PSF images of each science channel taken in the lab-test with the single and the multi time-step reconstructors. The color scale is linear and aligned for each channel. The wavelength is 1650 nm (H-band).}
\label{fig:psf}
\end{figure*}

We present the result of the laboratory experiment of the multi time-step reconstruction in \fref{fig:psf} and the top table of \tref{tab:labEE}. \fref{fig:psf} shows the PSF images of two science channels taken by the RAVEN infrared camera with an MOAO correction. The ensquared energy (EE) values included in a 140 mas box and $k_\text{SR}$ computed from the PSF images are summarized in \tref{tab:labEE}. The maximum wind speed simulated by the CU, which is 17.0 \ms, is almost half of the maximum wind speed in the model that we used for the numerical simulation in Section 3, and the time duration constraining the decay ratio $f$ of the temporal correlation of SH-WFS measurements to greater than 0.7 for a wind speed of 17.0 \ms\ is $\sim$100 ms. Therefore, we use 100 ms for $\Delta t$ of the multi time-step reconstructor. The pixel scale of the PSF image is 17.5 mas, and the size of a image is 0.5 arcsecond $\times$ 0.5 arcsecond. The performance of the MOAO is evaluated by SR, $k_\text{SR}$, and an EE at wavelength of 1650 nm (H-band), which is a ratio of intensity within a defined square to the total intensity. We use the simulated diffraction-limited PSF image to measure SR values of the PSF images taken in the lab-test. The lab-test PSF images are normalized by the total intensity within a 1 arcsecond box as same as the simulated PSF images. The size of a square for the EE is set to 140 mas, which is a size of the slit of the IRCS.\par
Visually, the PSFs become sharper and have higher peak intensity with the multi time-step reconstruction compared to the single time-step reconstruction for both of the asterisms, as shown in \Fref{fig:psf}. The EEs achieved by the single time-step reconstructor are $\sim$0.28 for the asterism without LGS and $\sim$0.35 with LGS. The improvement ratios of the EE achieved by the multi time-step reconstructor are 1.05--1.18. The absolute increases of the EE are 0.03--0.05. The SR is $\sim$0.08 with the single time-step reconstructor without the LGS, and this is consistent with the result of 0.10 measured in the laboratory experiment with the RAVEN CU previously \cite{Olivier-14}. On the other hand, the SR is $\sim$0.135 with the LGS, which is better than the one without LGS but lower than the previous value of 0.23 \cite{Olivier-14}. The improvement ratio, $k_\text{SR}$, is $\sim$1.4 without LGS and $\sim$1.23 with LGS. This result means the multi time-step reconstruction works well in the laboratory experiment. 
%
\section{DISCUSSION}
We compare the results of the laboratory experiment using the multi time-step reconstructor to a E2E numerical simulation. We assume the parameters of the RAVEN system and the turbulence profile generated by the RAVEN CU, listed in the top table of \tref{tab:slodar}. The GS and the target coordinates are also set to the same as the coordinates used in the lab-test, illustrated in \fref{fig:asterism}. The EE values and the improvement ratio, $k_\text{SR}$, predicted by the simulation are summarized in the lower part of \tref{tab:labEE}. There is a good agreement between the $k_\text{SR}$ measured in the lab-test and derived by the simulation, therefore the performance improvement of the multi time-step reconstructor is as predicted for both asterisms with and without LGS.\par
However, the EE and the SR measured in the lab-test are much smaller than the values predicted by the simulation for both of the single and the multi time-step reconstructors. This difference may be due to implementation errors not considered in the simulation, which are calibration error, DM control error, and an optical aberration. The improvement ratio, $k_\text{SR}$, is not affected by the implementation errors. Anderson et al. (2012) \cite{Andersen-12} simulates the performance of the RAVEN and estimates the error budget in the RAVEN system. They estimate a wavefront error (WFE) caused by the implementation errors of RAVEN, which includes calibration error, DM flattening, DM stability, DM repeatability, and high-order optical error, is 107 nm. The WFE due to the tomographic error is estimated from the SR derived from the simulation by using an approximation, SR=$\exp\left[-\sigma^2\right]$, where $\sigma=2\pi\text{WFE}/\lambda$ and $\lambda$=1650 nm. It is noted that this expression is valid for $\sigma^2<1$ rad$^2$, this is SR$<0.37$, and our result can be biased when SR is lower than 0.37. The total WFE is a quadrature sum of WFE due to the implementation errors and the tomographic error. We assume that the loss of the EE due to the implementation errors is the same as the loss of the SR. The performance predicted by simulation, accounting for the implementation errors, are listed in the bottom table of \tref{tab:labEE} as the values in a parenthesis. While the simulated EE including the implementation errors is slightly higher than the EE values measured in the lab-test, the simulated SR including the implementation error is still much higher than the measured SR. This suggests that there is an additional error, which makes a PSF peak blurred within a 140 mas box and affects mostly SR values. 
%
\section{CONCLUSIONS}
We propose the multi time-step reconstruction and a method estimating wind speeds and directions of multiple turbulence layers for tomographic WFAO systems. The multi time-step reconstruction is based on the frozen flow assumption and reduces the tomographic error caused by \textit{uncovered area}s and degeneracy due to \textit{unoverlapped area}s by increasing information of the atmospheric turbulence for the tomography by using the measurements at both of the previous and the current time-steps simultaneously. The wind estimation method estimates wind speed and direction at each altitude by using temporal correlation of the phase distortion pattern reconstructed by the tomographic reconstruction. \par
The numerical simulation, assuming an MOAO system on a 30 m aperture telescope with 8 LGSs at the top of Maunakea, shows that the multi time-step reconstructor with the time difference between the current and the previous time-step $\Delta t$=50 ms increases SR over a scientific FoV of 10 arcminutes diameter by a factor of 1.5$\sim$1.8, depending on guide star asterism, compared to the classical tomographic reconstructor if wind speeds and directions are given. Considering the time scale within which the frozen flow assumption is applicable, we set the time difference $\Delta t$ to keep the decay ratio of the peak intensity of temporal correlation to be more than 0.7. This is $\Delta t$=54 ms for the fastest layer with 33 \ms wind in the atmospheric model that we use for the simulation. The acceptable ranges for the wind estimation error are $-$7 \ms\ $\leq e_\perp\leq$ 7 \ms\ for the \textit{wind direction error}, $e_\perp$, and $-$5.5 \ms\ $\leq e_\parallel\leq$ 8 \ms\ for the \textit{wind speed error}.\par
We test the multi time-step reconstruction and the wind estimation method in the lab-test with the RAVEN CU. The wind speeds and directions at multiple layers are estimated successfully in the lab-test with the estimation error less than 2 \ms. The improvement ratio of EE thanks to the multi time-step reconstruction is 1.05--1.18 and the improvement ratio of SR is 1.17--1.44 depending on the GS asterism. This improvement is consistent with the prediction from a numerical simulation if we include the WFE due to the implementation of the AO system.\par
The multi time-step reconstructor will bring a potentially significant gain for MOAO instruments designed for the future ELTs, MOSAIC for European ELT \cite{Hammer-02} and TMT-AGE for TMT \cite{Akiyama-14,Ono-14}. The MOSAIC will use 6 LGSs at 7.5 arcminutes from the axis, and up to 5 additional NGSs inside of that field, depending upon availability. The TMT-AGE will have 8 LGSs within a scientific FoV of 10 arcminute diameter. With these large LGS separations, the \textit{uncovered} and \textit{unoverlapped area} might affects significantly on the final performance, even though the telescope aperture is large.\par
The main assumption for the multi time-step reconstruction is the frozen flow. The results in the analytical evaluation, the E2E simulation, and the laboratory experiment with RAVEN are computed with the perfect frozen flow condition. Although we refer to the decay ratio of the temporal-correlation of measurements from SH-WFSs as an indicator whether the frozen flow assumption is valid, it is still not clear how the boiling of the turbulence affects on the performance of the multi time-step reconstruction. In order to understand the effect of the boiling of the turbulence and the limitation of the time difference of the previous time-steps, on-sky experiments are needed. The computational complexity of the multi time-step reconstructor for ELT-sized systems is one of the points that must be considered. For instance, GPGPU can be used for accelerating the computation, and the predictive control can relax the temporal constraints.
%
%

\section*{Acknowledgments}
This work is supported by JSPS Grant-in-Aid for JSPS Fellows (15J02510). M. Akiyama is supported by JSPS Grant-in-Aid for Young Scientist (B) (23740140) and Grant-in-Aid for Scientific Research (B) (26287027). C. Correia acknowledges the support of the A*MIDEX project (n° ANR-11-IDEX-0001- 02) funded by the « Investissements d’Avenir » French Government program, managed by the French National Research Agency (ANR). Thanks to Naruhisha Takato, Yutaka Hayano, and Jean-Pierre V{\'e}ran for many discussions. Thanks to staff members of Subaru telescope for their support.

\appendix
\def\thesection{Appendix\Alph{section}}
\setcounter{equation}{0}
\def\theequation{\Alph{section}\arabic{equation}}
\section{Analytical tomographic error}
The tip-tilt removed tomographic error in the direction $\bm{\theta_k}$, $\sigma^2_\text{tomo}(\bm{\theta_k})$, is defined as
\begin{equation}\label{eqA1}
\sigma^2_\text{tomo}(\bm{\theta_k})=\langle ||\bm{\varphi_k}-\bm{\hat{\varphi}_k}||^2\rangle/n_\text{node},
\end{equation}
where $n_\text{node}$ is the number of valid nodes on the aperture-plane, $\bm{\varphi_k}$ and  $\bm{\hat{\varphi}_k}$ are an actual and a reconstructed aperture-plane phase coming from the direction $\bm{\theta_k}$, respectively, and $\langle\rangle$ indicates an ensemble average over time. With \eref{eq2A-1}, the actual phase is written as $\bm{\varphi_k}=\bm{P_{\theta_k}}\bm{\phi}$. The reconstructed phase is expressed using \eref{eq2A-3} and \eref{eq2A-4} as $\bm{\hat{\varphi}_k}=\bm{P_{\theta_k}}\bm{E}\bm{\Gamma}\bm{P_{gs}}\bm{\phi}+\bm{P_{\theta_k}}\bm{E}\bm{\eta}$, where $\bm{E}$ is a reconstructor reconstructing the phase distortion due to each atmospheric turbulence layer. Using these expressions, \eref{eqA1} is represented as 
\begin{equation}\label{eqA2}
\sigma^2_\text{tomo}(\bm{\theta_k})=\langle||\bm{P_{\theta_k}}(\bm{I}-\bm{E}\bm{\Gamma}\bm{P_{gs}})\bm{\phi}-\bm{P_{\theta_k}}\bm{E}\bm{\eta}||^2\rangle/n_\text{node},
\end{equation}
Since a variance of $\bm{x}$ can be rewritten with using a trace of a matrix as $\langle||\bm{x}||^2\rangle=\textbf{Tr}[\langle\bm{x}\bm{x}^T\rangle]$, we can get \eref{eq2C-1}, where we define that $\bm{\Sigma_\phi}=\langle\bm{\phi}\bm{\phi}^T\rangle$, $\bm{\Sigma_\eta}=\langle\bm{\eta}\bm{\eta}^T\rangle$, and it is assumed that the phase distortion $\bm{\phi}$ is independent from the measurement noise $\bm{\eta}$, $\langle\bm{\phi}\bm{\eta}^T\rangle=\langle\bm{\eta}\bm{\phi}^T\rangle=\bm{0}$.


\begin{table}[p]
\centering
\begin{tabular}{cc}
\hline
Parameters & Values \\
\hline
\hline 
Diameter of aperture & 30 m \\ 
scientific FoV  & 10 arcminutes\\
Zenith angle & $0^\circ$\\
Number of LGSs & 8 \\ 
Height of LGSs & 90 km \\
Number of turbulence layers & 7\\
$r_0$ at 500 nm & 0.156 m\\
$L_0$ & 30 m\\
WFS & SH-WFS \\
Number of WFS subapertures & 60$\times$60\\
Number of DM elements & 60$\times$60\\
Frame rate & 800 Hz \\
Time lag & 2 frames \\ 
count per a subaperture & 700 electrons \\
readout noise & 3 electrons \\
evaluate wavelength & 1650 nm (H-band)\\
\hline
\end{tabular}
\caption{Parameters used in the numerical simulation.}
\label{tab:setting}
\end{table}

\begin{table*}[p]
\centering
\begin{tabular}{ccccc}
\hline 
Altitude & $\int C_N^2 dh$ & Fraction & wind speed & wind direction \\
{[km]} & [$10^{-14}$ m$^{1/3}$]& of $C_N^2$ & [\ms] & [degree] \\
\hline 
16 & 2.734 & 0.0826 & 7.0   & 0\\
8   & 2.264 & 0.0684 & 33.0 & 45\\ 
4   & 2.879 & 0.0869 & 19.7 & 90\\
2   & 1.233 & 0.0372 & 11.6 & 135\\
1   & 1.074 & 0.0325 & 9.0  & 180\\
0.5& 3.190 & 0.0963 & 8.0   & 225\\
0   & 19.737 & 0.5960 & 7.0   & 270\\
\hline
Total $r_0$ [m] & 0.156 & & & \\
\hline
\end{tabular}
\caption{Atmospheric turbulence profile used in the numerical simulation.}
\label{tab:atmos_simu}
\end{table*}

\begin{table*}[p]
\centering
\begin{tabular}{ccccc}
\multicolumn{5}{c}{\textbf{Expected Profile}}\\
\hline 
Altitude & $\int C_N^2 dh$ & Fraction of& \multicolumn{2}{c}{wind speed [\ms]} \\
{[}km{]} & [$10^{-14}$ m$^{1/3}$] & $C_N^2$ & $x$ & $y$ \\
\hline 
10.5 & 5.961 & 0.180  & 17.0  & 0.0\\
5.5   & 7.418 & 0.224  & 6.0    & 0.0\\ 
0   & 19.737 & 0.596  & 0.0    & 5.68\\
\hline
Total $r_0$[m] & 0.156 & & & \\
\hline 
\end{tabular}\\
\vspace{0.7cm}
\begin{tabular}{ccccc}
\multicolumn{5}{c}{\textbf{Estimated Profile}}\\
\hline 
Altitude & $\int C_N^2 dh$ & Fraction of& \multicolumn{2}{c}{wind speed [\ms]} \\
{[}km{]} & [$10^{-14}$ m$^{1/3}$] & $C_N^2$ & $x$ & $y$ \\
\hline 
10 & 4.329 & 0.169  & 15.4  & -0.6\\
8   & 0.896 & 0.035 & 8.6 & -0.2\\ 
6   & 8.222 & 0.321  & 6.3 & 0.1\\ 
0   & 12.140 & 0.474  & 0.1 & 5.0\\
\hline
Total $r_0$[m] & 0.182 & & & \\
\hline 
\end{tabular}
\caption{Atmospheric turbulence profile and wind profile generated by the RAVEN CU (top) and estimated by the SLODAR and the wind estimation method (bottom).}
\label{tab:slodar}
\end{table*}

\begin{table*}[p]
\centering
\begin{tabular}{cccccc}
 \multicolumn{2}{c}{Lab-test}  & \multicolumn{2}{c}{EE (140mas)} & \multicolumn{2}{c}{SR}\\
 \hline
 reconstructor & GSs & Ch1 & Ch2 & Ch1 & Ch2\\
\hline 
\hline
single &\multirow{2}{*}{3NGSs}& 0.268 & 0.290  & 0.072 & 0.087\\
multi  & & 0.313 & 0.343  & 0.095 & 0.131\\
\hline
\multicolumn{2}{c}{Improvement ratio} & 1.16 & 1.18 & 1.32 & 1.51 \\
\hline
\hline 
single &3NGSs & 0.349 & 0.365 & 0.125 & 0.150\\
multi  & +LGS & 0.367 & 0.385 & 0.153 & 0.187\\
\hline
\multicolumn{2}{c}{Improvement ratio} &1.05 &1.05 & 1.22 & 1.25 \\
\hline
& & & & & \\
 \multicolumn{2}{c}{Simulation}  & \multicolumn{2}{c}{EE (140mas)} & \multicolumn{2}{c}{SR}\\
 \hline
 reconstructor & GSs & Ch1 & Ch2 & Ch1 & Ch2\\
\hline 
\hline
single & \multirow{2}{*}{3NGSs}& 0.334(0.294) & 0.346(0.303) & 0.251(0.211) & 0.267(0.224) \\
multi  & & 0.406(0.348) & 0.424(0.362) & 0.361(0.302) & 0.384(0.321)\\
\hline
\multicolumn{2}{c}{Improvement ratio} & 1.21(1.18) & 1.23(1.19) & 1.44(1.43) & 1.44(1.43) \\
\hline
\hline 
single & 3NGSs& 0.443(0.379) & 0.430(0.367) & 0.398(0.333) & 0.386(0.324)\\
multi  & +LGS & 0.477(0.402) & 0.477(0.402) & 0.465(0.390) & 0.465(0.389)\\
\hline
\multicolumn{2}{c}{Improvement ratio} & 1.07(1.06) & 1.11(1.10) & 1.17(1.17) & 1.20(1.20) \\
\hline
\end{tabular}
\caption{SR, $k_\text{SR}$, and EE value of each channel defined by a 140 mas box measured in the lab-test (the top table) compared to those predicted from the numerical simulation (the bottom table). There are two GS configuration used in the test, one is only three NGSs and the other is three NGSs+LGS. The values in parenthesis in the lower part is the value accounting for the implementation errors.}
\label{tab:labEE}
\end{table*}


\begin{thebibliography}{10}
\newcommand{\enquote}[1]{``#1''}

\bibitem{Hayano-10}
Y.~Hayano, H.~Takami, S.~Oya, M.~Hattori, Y.~Saito, M.~Watanabe, O.~Guyon,
  Y.~Minowa, S.~E. Egner, M.~Ito, V.~Garrel, S.~Colley, T.~Golota, and M.~Iye,
  \enquote{Commissioning status of {Subaru} laser guide star adaptive optics
  system,} Proc. SPIE \textbf{7736}, 77360N (2010).

\bibitem{Wizinowich-06}
P.~L. Wizinowich, D.~L. Mignant, A.~H. Boucheza, R.~D. Campbell, J.~C.~Y. Chin,
  A.~R. Contos, M.~A. van Dam, S.~K. Hartman, E.~M. Johansson, R.~E. Lafon,
  H.~Lewis, P.~J. Stomski, D.~M. Summers, C.~G. Brown, P.~M. Danforth, C.~E.
  Max, and D.~M. Pennington, \enquote{The w. m. keck observatory laser guide
  star adaptive optics system: Overview,} Publications of the Astronomical
  Society of the Pacific \textbf{118}, 297--309 (2006).

\bibitem{Lai-14}
O.~Lai, J.-P. Véran, G.~Herriot, J.~White, J.~Ball, and C.~Trujillo,
  \enquote{Altair performance and upgrades,} Proc. SPIE \textbf{9148},
  914838--914838--13 (2014).

\bibitem{Skidmore-09}
W.~Skidmore, S.~Els, T.~Travouillon, R.~Riddle, M.~Schöck, E.~Bustos,
  J.~Seguel, and D.~Walker, \enquote{Thirty meter telescope site testing v:
  Seeing and isoplanatic angle,} Publications of the Astronomical Society of
  the Pacific \textbf{121}, 1151--1166 (2009).

\bibitem{Beckers-88}
J.~Beckers, \enquote{Increasing the size of the anisoplanatic patch with
  multiconjugate adaptive optics,} Proc. ESO conference \textbf{4007}, 693--703
  (1988).

\bibitem{Hammer-02}
F.~Hammer, F.~Sayède, E.~Gendron, T.~Fusco, D.~Burgarella, V.~Cayatte,
  J.~Conan, F.~Courbin, H.~Flores, I.~Guinouard, L.~Jocou, A.~Lançon,
  G.~Monnet, M.~Mouhcine, F.~Rigaud, D.~Rouan, G.~Rousset, V.~Buat, and
  F.~Zamkotsian, \enquote{The falcon concept: multi-object spectroscopy
  combined with mcao in near-ir,} Proc. ESO workshop p. 139 (2002).

\bibitem{Andersen-06}
D.~R. Andersen, S.~S. Eikenberry, M.~Fletcher, W.~Gardhouse, B.~Leckie, J.-P.
  Véran, D.~Gavel, R.~Clare, R.~Guzman, L.~Jolissaint, R.~Julian, and
  W.~Rambold, \enquote{The moao system of the irmos near-infrared multi-object
  spectrograph for tmt,} Proc. SPIE \textbf{6269}, 62694K--62694K--12 (2006).

\bibitem{Hammer-14}
F.~Hammer, B.~Barbuy, J.~G. Cuby, L.~Kaper, S.~Morris, C.~J. Evans,
  P.~Jagourel, G.~Dalton, P.~Rees, M.~Puech, M.~Rodrigues, D.~Pearson, and
  K.~Disseau, \enquote{Mosaic at the e-elt: A multi-object spectrograph for
  astrophysics, igm and cosmology,} Proc. SPIE \textbf{9147}, 914727--914727--7
  (2014).

\bibitem{Akiyama-14}
M.~Akiyama, S.~Oya, Y.~H. Ono, H.~Takami, S.~Ozaki, Y.~Hayano, I.~Iwata,
  K.~Hane, T.~Wu, T.~Yamamuro, and Y.~Ikeda, \enquote{Tmt-age: wide field of
  regard multi-object adaptive optics for tmt,} Proc. SPIE \textbf{9148},
  914814--914814--14 (2014).

\bibitem{Correia-14}
C.~Correia, K.~Jackson, J.-P. V\'{e}ran, D.~Andersen, O.~Lardi\`{e}re, and
  C.~Bradley, \enquote{Static and predictive tomographic reconstruction for
  wide-field multi-object adaptive optics systems,} J. Opt. Soc. Am. A
  \textbf{31}, 101--113 (2014).

\bibitem{Correia-15}
C.~M. Correia, K.~Jackson, J.-P. V\'{e}ran, D.~Andersen, O.~Lardi\`{e}re, and
  C.~Bradley, \enquote{Spatio-angular minimum-variance tomographic controller
  for multi-object adaptive-optics systems,} Appl. Opt. \textbf{54}, 5281--5290
  (2015).

\bibitem{Ammons-12}
S.~M. Ammons, L.~Poyneer, D.~T. Gavel, R.~Kupke, C.~E. Max, and L.~Johnson,
  \enquote{Evidence that wind prediction with multiple guide stars reduces
  tomographic errors and expands moao field of regard,} Proc. SPIE
  \textbf{8447}, 84471U (2012).

\bibitem{Poyneer-09}
L.~Poyneer, M.~van Dam, and J.-P. V\'{e}ran, \enquote{Experimental verification
  of the frozen flow atmospheric turbulence assumption with use of astronomical
  adaptive optics telemetry,} J. Opt. Soc. Am. A \textbf{26}, 833--846 (2009).

\bibitem{Guesalaga-14}
A.~Guesalaga, B.~Neichel, A.~Cortés, C.~Béchet, and D.~Guzmán,
  \enquote{Using the ${C_{n}^{2}}$ and wind profiler method with wide-field
  laser-guide-stars adaptive optics to quantify the frozen-flow decay,} Monthly
  Notices of the Royal Astronomical Society \textbf{440}, 1925--1933 (2014).

\bibitem{Olivier-14}
O.~Lardière, D.~Andersen, C.~Blain, C.~Bradley, D.~Gamroth, K.~Jackson,
  P.~Lach, R.~Nash, K.~Venn, J.-P. Véran, C.~Correia, S.~Oya, Y.~Hayano,
  H.~Terada, Y.~Ono, and M.~Akiyama, \enquote{{Multi-object adaptive optics
  on-sky results with Raven},} Proc. SPIE \textbf{9148}, 91481G (2014).

\bibitem{Ellerbroek-02}
B.~L. Ellerbroek, \enquote{Efficient computation of minimum-variance wave-front
  reconstructors with sparse matrix techniques,} J. Opt. Soc. Am. A
  \textbf{19}, 1803--1816 (2002).

\bibitem{Gilles-13}
L.~Gilles, P.~Massioni, C.~Kulcs\'{a}r, H.-F. Raynaud, and B.~Ellerbroek,
  \enquote{{Distributed} {Kalman} filtering compared to {Fourier} domain
  preconditioned conjugate gradient for laser guide star tomography on
  extremely large telescopes,} J. Opt. Soc. Am. A \textbf{30}, 898--909 (2013).

\bibitem{Lee-07}
L.~H. Lee, \enquote{Sparse-matrix regularization for
  minimum-variancereconstruction of pseudo-kolmogorov turbulence,} in
  \enquote{Adaptive Optics: Analysis and Methods/Computational Optical Sensing
  and Imaging/Information Photonics/Signal Recovery and Synthesis Topical
  Meetings on CD-ROM,}  (Optical Society of America, 2007), p. JTuA2.

\bibitem{Schock-00}
M.~Sch\"{o}ck and E.~J. Spillar, \enquote{Method for a quantitative
  investigation of the frozen flow hypothesis,} J. Opt. Soc. Am. A \textbf{17},
  1650--1658 (2000).

\bibitem{Gilles-02}
L.~Gilles, C.~R. Vogel, and B.~L. Ellerbroek, \enquote{Multigrid preconditioned
  conjugate-gradient method for large-scale wave-front reconstruction,} J. Opt.
  Soc. Am. A \textbf{19}, 1817--1822 (2002).

\bibitem{Yang-06}
Q.~Yang, C.~R. Vogel, and B.~L. Ellerbroek, \enquote{Fourier domain
  preconditioned conjugate gradient algorithm for atmospheric tomography,}
  Appl. Opt. \textbf{45}, 5281--5293 (2006).

\bibitem{Lessard-08}
L.~Lessard, M.~West, D.~MacMynowski, and S.~Lall, \enquote{Warm-started
  wavefront reconstruction for adaptive optics,} J. Opt. Soc. Am. A
  \textbf{25}, 1147--1155 (2008).

\bibitem{Gilles-06}
L.~Gilles and B.~Ellerbroek, \enquote{Shack-hartmann wavefront sensing with
  elongated sodium laser beacons: centroiding versus matched filtering,} Appl.
  Opt. \textbf{45}, 6568--6576 (2006).

\bibitem{Andersen-12}
D.~R. Andersen, K.~J. Jackson, C.~Blain, C.~Bradley, C.~Correia, M.~Ito,
  O.~Lardière, and J.-P. Véran, \enquote{Performance {Modeling} for the
  {RAVEN} {Multi-Object Adaptive Optics Demonstrator},} Publications of the
  Astronomical Society of the Pacific \textbf{124}, pp. 469--484 (2012).

\bibitem{Miyashita-04}
A.~Miyashita, N.~Takato, T.~Usuda, F.~Uraguchi, and R.~Ogasawara,
  \enquote{Statistics of the weather data, environment data, and the seeing of
  the {Subaru Telescope},} Proc. SPIE \textbf{5489}, 207--217 (2004).

\bibitem{Els-09}
S.~G. Els, T.~Travouillon, M.~Schöck, R.~Riddle, W.~Skidmore, J.~Seguel,
  E.~Bustos, and D.~Walker, \enquote{{Thirty Meter Telescope Site Testing VI:
  Turbulence Profiles},} Publications of the Astronomical Society of the
  Pacific \textbf{121}, pp. 527--543 (2009).

\bibitem{Hardy-98}
J.~W. Hardy, \emph{Adoptive Optics for Astronomical Telescopes} (Oxford
  University Press, 1998).

\bibitem{Wilson-02}
R.~W. Wilson, \enquote{{SLODAR}: measuring optical turbulence altitude with a
  {Shack–Hartmann} wavefront sensor,} Monthly Notices of the Royal
  Astronomical Society \textbf{337}, 103--108 (2002).

\bibitem{Butterley-06}
T.~Butterley, R.~W. Wilson, and M.~Sarazin, \enquote{Determination of the
  profile of atmospheric optical turbulence strength from {SLODAR} data,}
  Monthly Notices of the Royal Astronomical Society \textbf{369}, 835--845
  (2006).

\bibitem{Cortes-12}
A.~Cortés, B.~Neichel, A.~Guesalaga, J.~Osborn, F.~Rigaut, and D.~Guzman,
  \enquote{Atmospheric turbulence profiling using multiple laser star wavefront
  sensors,} Monthly Notices of the Royal Astronomical Society \textbf{427},
  2089--2099 (2012).

\bibitem{Lavigne-12}
J.-F. Lavigne, F.~Lamontagne, G.~Anctil, M.~Wang, M.~Tremblay, O.~Lardière,
  R.~Nash, D.~Andersen, M.~Savard, P.~Côté, C.~H. Bradley, and
  F.~Châteauneuf, \enquote{Design and test results of the calibration unit for
  the {MOAO} demonstrator {RAVEN},} Proc. SPIE \textbf{8447}, 844754 (2012).

\bibitem{Ono-14}
Y.~H. Ono, M.~Akiyama, and S.~Oya, \enquote{Tmt-age: numerical simulation of a
  new tomographic reconstruction method for wide for moao,} Proc. SPIE
  \textbf{9148}, 91486M--91486M--6 (2014).

\end{thebibliography}
\end{document}